\documentclass[lettersize,journal]{IEEEtran}
\usepackage{amsmath,amsfonts}
\usepackage{algorithmic}
\usepackage{algorithm}
\usepackage{array}
\usepackage[caption=false,font=normalsize,labelfont=sf,textfont=sf]{subfig}
\usepackage{textcomp}
\usepackage{stfloats}
\usepackage{url}
\usepackage{verbatim}
\usepackage{graphicx}
\usepackage{colortbl}
\usepackage{cite}
\usepackage{multirow}
\usepackage{amsmath}
\usepackage{booktabs}
\usepackage{enumitem}
\usepackage{color}
\usepackage{xspace}
\newcommand{\ie}{{\emph{i.e.}},\xspace}

\newcommand{\eg}{{\emph{e.g.}},\xspace}

\newcommand{\etal}{{\emph{et al.}}}

\usepackage{xcolor}
\definecolor{mydarkred}{rgb}{0.6,0,0}
\definecolor{mydarkgreen}{rgb}{0,0.6,0}
\definecolor{mydarkblue}{rgb}{0,0,0.6}
\usepackage[colorlinks,
linkcolor=mydarkred,
citecolor=mydarkgreen]{hyperref}

\newtheorem{theorem}{Theorem}[section]

\begin{document}
\title{Redundancy-Adaptive Multimodal Learning for Imperfect Data}

\author{Mengxi Chen, Jiangchao Yao, Linyu Xing, Yu Wang, Ya Zhang, Yanfeng Wang
        
\thanks{Mengxi Chen and Linyu Xing are with the Cooperative Medianet Innovation Center of Shanghai Jiao Tong University
(e-mail: \{mxchen\_mc, xly8991\}@sjtu.edu.cn).}
\thanks{Jiangchao Yao, Yu Wang, Ya Zhang and Yanfeng Wang are with the Cooperative Medianet Innovation Center of Shanghai Jiao Tong University and Shanghai AI Laboratory
(e-mail: \{Sunarker, yuwangsjtu, ya\_zhang, wangyanfeng622\}@sjtu.edu.cn).}
}
\markboth{Journal of \LaTeX\ Class Files,~Vol.~14, No.~8, August~2021}%
{Chen \MakeLowercase{\textit{et al.}}: Redundancy-Adaptive Multimodal Learning for Imperfect Data}


\maketitle

\begin{abstract}
Multimodal models trained on complete modality data often exhibit a substantial decrease in performance when faced with imperfect data containing corruptions or missing modalities. To address this robustness challenge, prior methods have explored various approaches from aspects of augmentation, consistency or uncertainty, but these approaches come with associated drawbacks related to data complexity, representation, and learning, potentially diminishing their overall effectiveness. In response to these challenges, this study introduces a novel approach known as the Redundancy-Adaptive Multimodal Learning (RAML). RAML efficiently harnesses information redundancy across multiple modalities to combat the issues posed by imperfect data while remaining compatible with the complete modality. Specifically, RAML achieves redundancy-lossless information extraction through separate unimodal discriminative tasks and enforces a proper norm constraint on each unimodal feature representation. Furthermore, RAML explicitly enhances multimodal fusion by leveraging fine-grained redundancy among unimodal features to learn correspondences between corrupted and untainted information. Extensive experiments on various benchmark datasets under diverse conditions have consistently demonstrated that RAML outperforms state-of-the-art methods by a significant margin.
\end{abstract}


\section{Introduction}

\begin{figure}[!t]
 \centering
 \includegraphics[width=0.99\linewidth]{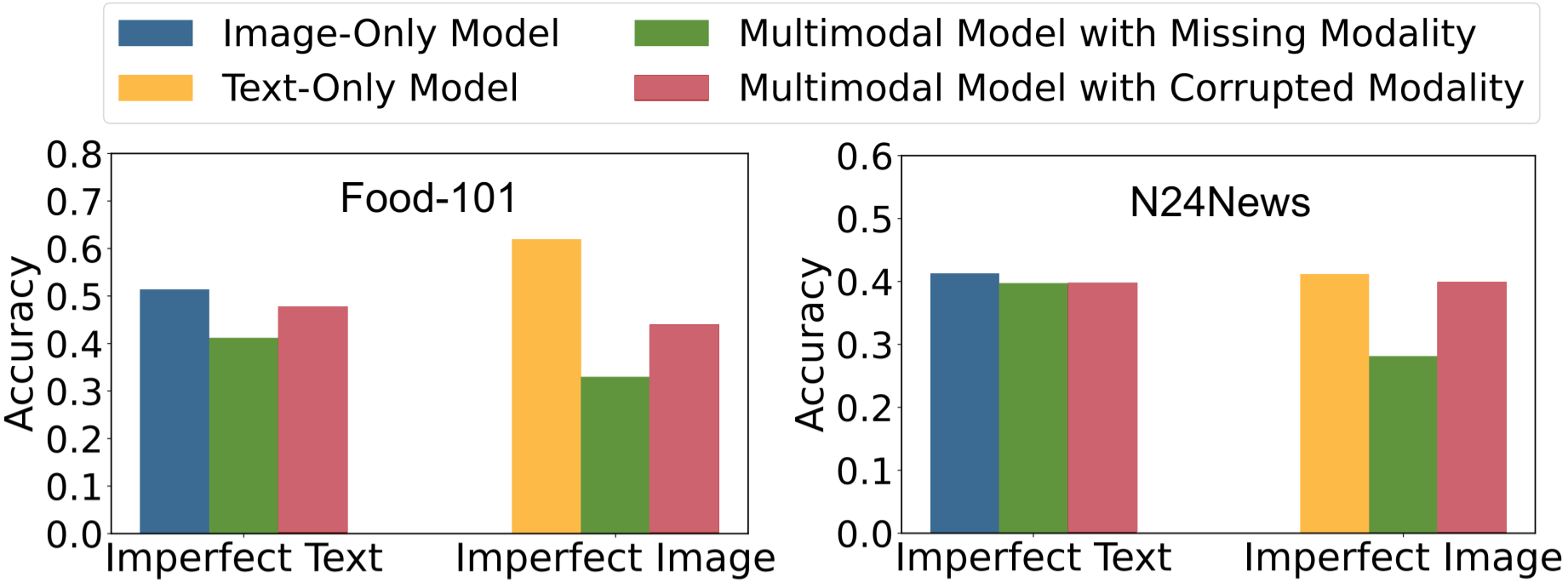}
    \caption{
    Comparison of the unimodal and multimodal models~\cite{DBLP:journals/tgrs/HongGYYCDZ21} on Food-101~\cite{bossard2014food} and N24News~\cite{wang2021n15news} when encountering imperfect data during testing. The multimodal model that concatenates unimodal features at the intermediate layer is trained on full-modality data while the unimodal models are trained on the corresponding unimodality data. As can be seen, when the multimodal model is fed with imperfect data, it can be even worse than unimodal models. 
    }
    \label{fig1}
\end{figure}

Complementary information offered by the combination of multiple modalities~\cite{huang2021makes,springstein2021quti,wang2020makes,Huang2021WhatMM} are usually expected to augment the effectiveness of multimodal applications~\cite{abdu2021multimodal,DBLP:conf/emnlp/HanCP21,zadeh2018multimodal,chen2019robust,zhang2018multimodal}.  
However, the success of such approaches hinges critically on the assumption that all data remains uncorrupted and complete across modalities during both training and testing, a presumption that often does not align with real-world scenarios.
For instance, in emotion recognition, the absence of a modality—be it vision, audio, or text—can arise due to factors such as camera non-coverage, ambient noise, or privacy concerns~\cite{zeng2022tag}. 
When data modalities are less than perfect, the performance of multimodal models trained on flawless samples deteriorates significantly. As depicted in Figure~\ref{fig1}, it may even fall below the performance of an unimodal model. This phenomenon can be attributed to the neural network's inability to concurrently represent redundant information across different modalities.
Consequently, when modalities are imperfect, previous multimodal models fail to harness the redundant information initially captured from these modalities, even if such information remains untainted in other modalities. This inevitably leads to a degradation in performance.


While capturing complementary information from each unimodality enhances performance, the key to improving multimodal robustness lies in effectively exploiting redundancy among modalities.
One straightforward approach is to simulate imperfect data through data augmentation, allowing for full consideration of modality redundancy~\cite{bednarek2020robustness, kim2019robust}.
However, brute-force data augmentation can exacerbate training difficulties, especially when dealing with sophisticated datasets~\cite{hao2023mixgen}.
Recent consistency-based methods, which focus on learning the consistency among modalities, enhance robustness by reconstructing data from imperfect modalities~\cite{shi2019variational,tran2017missing,wu2018multimodal,9258396,xaviar2023robust}, or by introducing cycle-consistency constraints on unimodal representations~\cite{zeng2022tag,zhao2021missing}. 
Nevertheless, these approaches tend to overly prioritize achieving better alignment among modalities in order to learn redundancy, often overlooking the importance of complementary information.
Other methods employ uncertainty estimation to weigh each unimodal representation, thereby mitigating the influence of imperfect data on the final decision~\cite{han2022multimodal,DBLP:conf/iclr/HanZFZ21,DBLP:journals/pami/HanZFZ23}. However, this globally weakened contribution of imperfect data can result in the underestimation of its valuable untainted information.

\begin{figure}[tp]
    \centering
    \includegraphics[width = 0.99\linewidth]{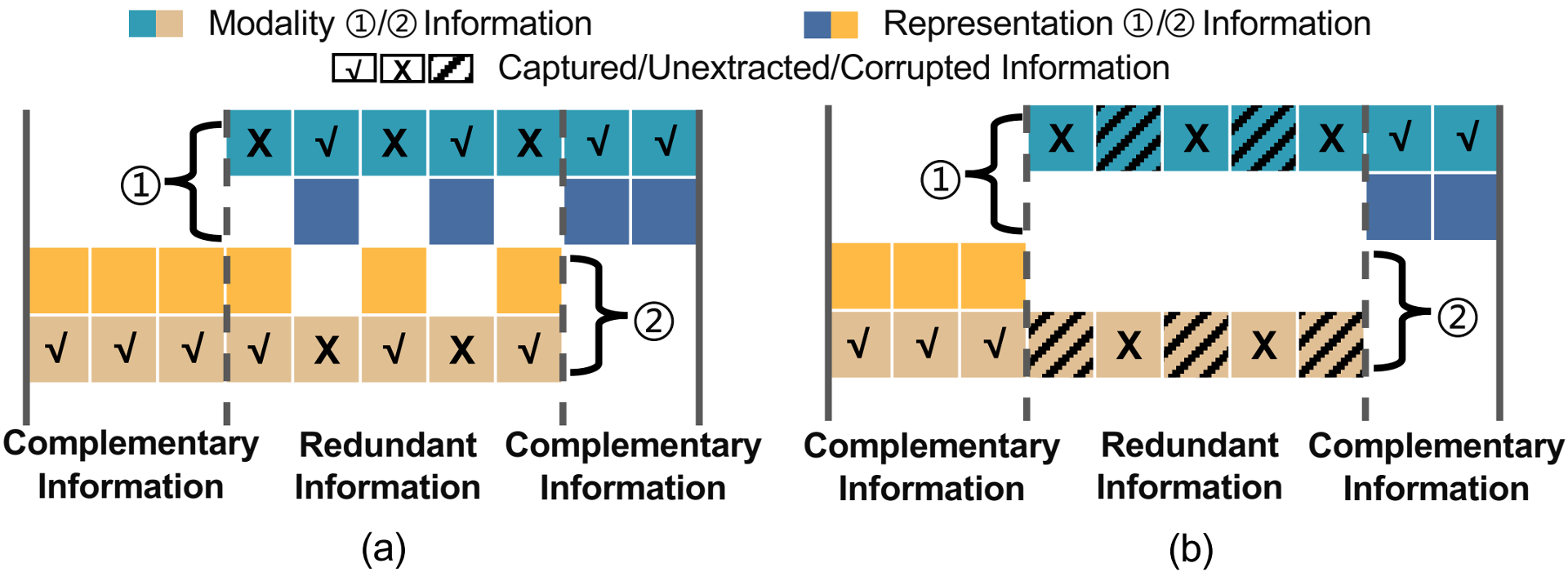}
    \caption{(a) The redundant information is partially expressed in representations of different modalities. (b) When the redundant information suffers from corruption, the untainted redundant counterpart is not used for remedy.
    }
    \label{fig1-2}
\end{figure}

Figure~\ref{fig1-2}(a) illustrates an example under the condition of perfect data training, where different unimodal representations extract only partial redundant information, alongside their own complementary information. 
It is evidence-sufficient for prediction and performs admirably when the data remains perfect during testing. 
However, in the contrasting scenario depicted in Figure~\ref{fig1-2}(b), where the captured redundant information is corrupted, the multimodal model fails to tap into the remaining untainted redundant information from other modalities for recovery (as indicated by the $\times$ marked boxes).
Thus, it becomes imperative to fully consider data redundancy across all modalities to achieve multimodal robustness.
Intuitively, this consideration should encompass two critical aspects: 1) learning a feature representation for each modality that comprehensively captures task-relevant information, and 2) assessing the quality of the feature representation in a fine-grained manner to gauge its correspondence between both corrupted and untainted information. This fine-grained assessment is crucial for facilitating optimal fusion and, consequently, enhancing multimodal robustness.

Motivated by the above analysis, we propose a novel method named Redundancy-Adaptive Multimodal Learning (RAML) to address the complexities of imperfect multimodal data. 
To begin, we implement a unique strategy to address the first point: We parameterize each unimodal feature representation as a probabilistic distribution through distinct discriminative modeling. Besides, we apply a judicious sparse constraint to these features, inherently fostering the lossless extraction of all pertinent information, including both redundant and complementary components, while guarding against overfitting.
As for the second point, the primary challenge centers around effectively perceiving and aggregating the pristine information from each modality in a high-quality manner. To tackle this, we introduce an explicit weighting mechanism that adaptively gauges the element-wise information quality of each unimodal feature representation through the variance analysis, thereby maximizing the utilization of untainted information.
In a nutshell, our contributions can be summarized as follows:
\begin{itemize}
\item We identify the key to multimodal robustness and propose a novel Redundancy-Adaptive Multimodal Learning method that efficiently utilizes the redundancy among modalities to combat the challenge of imperfect data.

\item We propose to construct separate unimodal discriminative tasks with the sparsity constraint to maximally preserve the unimodal redundant and complementary information. Additionally, the designed element-wise feature quality estimation makes it possible to aggregate the untainted information of each modality effectively.

\item We compare the theoretical upper bound of the predictive ground-truth label probability of the weighting mechanism in RAML with the bounds of uniform and coarse-grained weighting mechanisms, which shows the superiority of the fine-grained nature in RAML.

\item We conduct extensive experiments to show the promise of RAML under different conditions, \eg full modalities, corrupted modalities, and missing modalities. The experimental results confirm the consistent superiority of our proposed method under different conditions.

\end{itemize}

\section{Related Work}
In the following, we retrospect the early study for multimodal robustness in the perspectives of data augmentation, consistency-based methods and uncertainty-based learning.

\subsection{Data Augmentation-based Robustness}
Data augmentation is a common technique used to enhance multimodal robustness by adding noise and other modifications to simulate imperfect data, which implicitly promotes the learning of redundancy. Bednarek \etal~\cite{bednarek2020robustness} showed that each dataset often has a dominant modality, and proposed adding noise or setting zeros to the dominant modality of training samples to reduce the negative impact of input quality degradation.  Zhao \etal~\cite{zhao-calapodescu-2022-multimodal} manipulated different noise types on the training samples and insert noise adapters in the network to specifically handle the different types of noise. Kim \etal~\cite{kim2019robust} used techniques such as blanking, noise addition, and occlusion to generate additional examples and guide the network to learn to fuse features appropriately in adverse environments. However, the strengthened data augmentation also exacerbates the training difficulty, which conversely makes the multimodal information integration more difficult.

\subsection{Consistency-based Multimodal Robustness} 

This line of works target to handle the missing-modality issue in multimodal learning~\cite{hoffman2016learning,hu2020knowledge,poklukar2022geometric,chen2023enhanced,zhang2021Modality,Zhang2022mmFormer} by training a collection of models for any combination of modalities, which are mainly limited by the expensive training and deployment cost with the exponentially increasing parameters. Recent efforts to address this problem are from the aspects of reconstruction~\cite{shi2019variational,tran2017missing,wu2018multimodal,9258396} or joint learning~\cite{DBLP:conf/cvpr/0002R0T022,zeng2022tag,zhao2021missing} based on the consistency among modalities.  The former learns to reconstruct the missing modalities via the observed modalities. For example, MVAE~\cite{wu2018multimodal} used a product-of-experts (POE) inference network and a sub-sampled training paradigm to solve the missing modality inference problem. MMVAE~\cite{shi2019variational} learned the robust joint variational posterior through the mixture of each single-modality posterior.
The latter learns joint representations by enforcing the modality consistency. For instance, 
MMIN~\cite{zhao2021missing} proposed a unified multimodal recognition model that adopts the cascade residual AE and the cycle consistency learning to impute the missing-modality embedding. TATE~\cite{zeng2022tag} constructed a new common space projection module with the missing modality tag encoding for robust sentiment analysis. Both reconstruction and joint representation learning  emphasize the alignment among modalities but ignore the modality-complementary part, which limits the overall improvement in robust multimodal learning. 

\begin{figure*}[ht!]
    \centering
    \includegraphics[width = 0.98\linewidth]{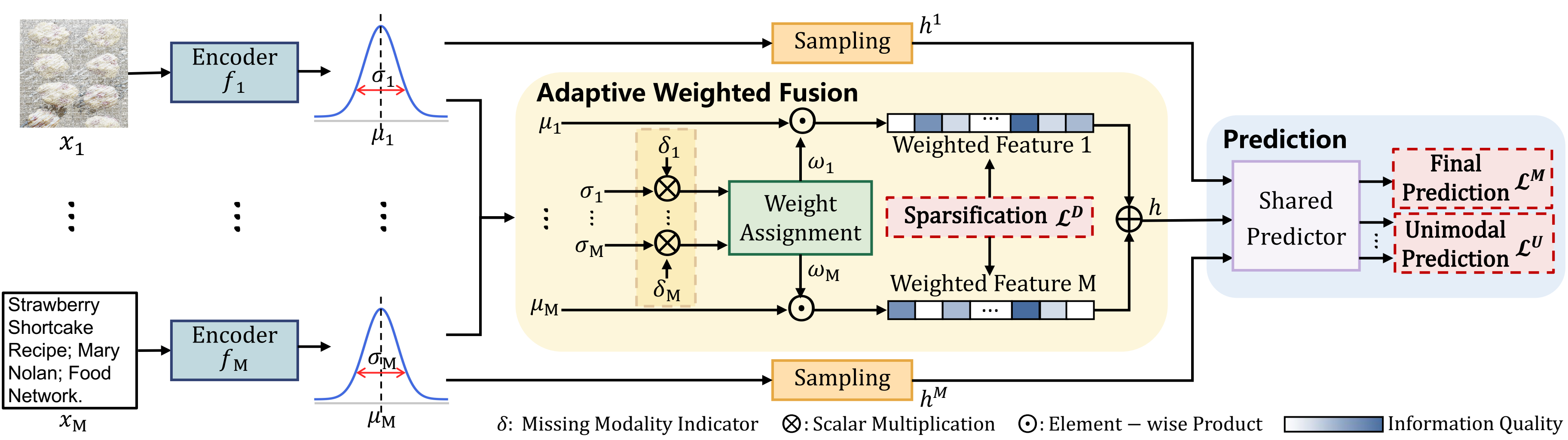}
    \caption{An overview of RAML. Each modality branch has one individual encoder to model the corresponding unimodal representation as a normal distribution.
    An explicit weighting mechanism is constructed to measure the element-wise information quality of each unimodal representation, which promotes the multimodal fusion process.
    The separate unimodal and multimodal discriminative tasks with a shared predictor and the sparse constraint on each unimodal representation are used to capture robust informative features for each modality. 
    }
    \label{fig2}
\end{figure*}

\subsection{Uncertainty-based Multimodal Robustness} 
The study in this direction is to construct the robust fusion of modalities via uncertainty modeling.
For example, Subedar \etal~\cite{subedar2019uncertainty} applied Bayesian Deep Neural Networks for uncertainty-aware audiovisual fusion to improve the human activity recognition. 
TMC~\cite{DBLP:conf/iclr/HanZFZ21} used the Dirichlet distribution to model the uncertainties of multi-view data without introducing additional parameters and used Dempster's Rule to fuse multimodal features. 
The subsequent ETMC~\cite{DBLP:journals/pami/HanZFZ23} introduced a pseudo-view to promote interaction between different views, while maintaining the trustworthiness of the original TMC.
Recently, MD~\cite{han2022multimodal} employed the true-class probability to obtain reliable unimodal confidence and used a weighting strategy to incorporate the modality information. However, although the uncertainty-based methods can fuse modalities by quality, most of them consider the coarse fusion of full modalities, which is not robust to the imperfect data.

\section{Method}
\subsection{Preliminary}
Suppose that we have a training set of $\big\{\left(x^i, y^i\right)\big\}_{i=1}^N$, where $x^i$ is an input with $M$ modalities, namely, $x^i=\{x^i_m\}_{m=1}^M$, and $y^i \in \{1,\cdots,C\}$ is the corresponding label. $N$ and $C$ are the size of the dataset and the class number respectively. Our goal is to predict the label $y^i$ under imperfect $x^i$. In this work, we consider $x^i$ under the modality missing or corruption conditions. Specially, regarding the missing modalities, there are $2^M-1$ meaningful combinations that have excluded the fully-missing case. If one modality is missing, the corresponding unimodal data is replaced by zero. We introduce an indicator set $\delta^i=\{\delta^i_m\}^M_{m=1}$, where $\delta^i_m\in\{0,~1\}$, to indicate whether the modality $m$ of $x^i$ is missing or not.
During both training and testing, we augment full-modality samples $x^i$ with all possible missing patterns to simulate the missing modality conditions.
Regarding the corruption, we have $\forall m,~\delta^i_m=1$, while some of the modalities can be corrupted by noise, and we will manually add the noise of different strengths into the original $x^i$ during testing.

\subsection{The Proposed Method}
To handle imperfect multimodal data, the proposed method designs redundancy-lossless unimodal learning for fully capturing the useful redundant and complementary information (Section~\ref{sec3.2}) and redundancy-adaptive multimodal fusion for perceiving and integrating the untainted information (Section~\ref{sec3.3}). We summarize the overall loss in Section~\ref{sec3.2.3} and illustrate the framework of RAML in Figure~\ref{fig2}.

\subsubsection{\textbf{Redundancy-Lossless Unimodal Learning}}
\label{sec3.2}
Given a multimodal sample $x^i$, to guarantee the sufficient information extraction of unimodalities, we construct individual model-specific encoders for each of them to learn their corresponding representations. Specifically, we parameterize the representation of each modality as a $D$-dimensional normal distribution, whose mean and the covariance are directly modeled by the modality-specific encoder as follows:
\begin{align}
\begin{split}
    &q\left(h^i_m|x^i_m\right) \sim \mathcal{N}\left(\mu^i_m,(\sigma^i_m)^2\right),\\
    &\mu^i_m=g^{\mu}_m \left(f_m\left(x^i_m\right)\right), \, \sigma^i_m=g^{\sigma}_m \left(f_m\left(x^i_m\right)\right),
\end{split}
\end{align}
where $\mu_m^i,\sigma_m^i \in \mathbb{R}^D$ are the mean and variance vectors of the normal distribution $\mathcal{N}(\mu^i_m,\sigma^i_m)$ respectively, $f_m(\cdot)$ is the encoder of the $m$-th modality, and $g^{\mu}_m(\cdot)$, $g^{\sigma}_m(\cdot)$ are two head modules for computing the mean and variance vectors. Note that, the reason that we adopt the probabilistic modeling for the representation of each modality is twofold: 1) the probabilistic modeling encodes a distribution of possible values for each unimodal representation, rather than a single deterministic vector, which makes it more tolerant to small perturbations in data compared to the deterministic modeling; 2) the variance of the probabilistic distribution provides us with a chance to estimate the element-wise quality of each unimodal representation, which is important for identifying the untainted information in the subsequent fusion.

\noindent\emph{\textbf{Learning Lossless Representation by Classification.}} As discussed in the previous section, how the redundant information among modalities can be learned intrinsically limit the robustness of multimodal learning. To fully capture the information useful for the task without ignoring either redundant or complementary parts among modalities, we construct a discriminative task to supervise the representation learning of each modality, which can be formulated as follows:
\begin{align}
    \mathcal{L}^U = \sum_{m=1}^M  \mathcal{\ell}\left(y^i, f\left(h^i_m\right)\right)\bigg|_{h_m^i\sim q(h^i_m|x^i_m)},
    \label{eq:cls^S}
\end{align}
where $f(\cdot)$ is the modality-shared predictor that takes each sampled representation $h^i_m$ from $\mathcal{N}\left(\mu^i_m,\sigma^i_m\right)$ as input and generates the prediction. $\ell$ is the cross-entropy loss. The shared predictor used in Eq.~(\ref{eq:cls^S}) maps the unimodal embeddings into the same latent space, making those embeddings comparable. About training with the probabilistic variable, we apply reparameterization tricks~\cite{kingma2013auto} to ease the gradient backpropagation, where $h^i_m=\mu^i_m + \rho^i_m \sigma^i_m$ under perturbation $\rho^i_m\sim\mathcal{N}(0,1)$.

\noindent\emph{\textbf{Eliminating Spurious Extraction via Sparsification.}} As the single modality might not contain the totally complete information causally associated with the label, arbitrarily optimizing Eq.~\eqref{eq:cls^S} for each modality will induce the spurious information extraction.
To address this dilemma, we propose to constrain the information extraction of $f_m(\cdot)$ from the perspective of the sparsification. As we know, it is effective to guarantee the proper sparsity by minimizing $l_1$ norm to avoid overfitting~\cite{bishop2006pattern}. We can similarly apply the $l_1$ norm constraint on our unimodal representation learning as follows,
\begin{align}
\begin{split}
    &\mathcal{L}^{D} = \sum_{m=1}^M \big|\big|\left(\frac{\sigma^i}{\sigma^i_m}\right)^2 \odot \mu^i_m\big|\big|_1, \; \\
    &\mathrm{where} \; \frac{1}{\left(\sigma^i\right)^2}= \sum_{m=1}^M \frac{1}{\left(\sigma_m^i\right)^2}.
        \label{eq5}
\end{split}
\end{align}
Note that, different from the sparsity on the single vector, we propose to constrain the magnitude of both mean and variance simultaneously. In our probabilistic modeling, the elements with large variances are considered unreliable while a small mean value indicates the element contains less information. 
Eq.(\ref{eq5}) promotes the elimination of the uninformative unimodal contents $\big|\big|(\frac{\sigma^i}{\sigma^i_m})^2 \odot \mu^i_m\big|\big|_1$ in two folds, namely, making $\mu$ as small as possible and making $\sigma$ as large as possible.
In total, the combination of Eq.(\ref{eq:cls^S}) and Eq.(\ref{eq5}) helps each unimodal encoder capture robust informative features for each modality.

\subsubsection{\textbf{Redundancy-Adaptive Multimodal Fusion}}
\label{sec3.3}

As discussed in the previous section, a better fusion builds based on the accurate quality measure about representation, especially, in a fine-grained nature to reflect its correspondence to the corrupt and untainted information. To achieve this goal, we leverage the variance analysis of unimodal representation to construct an explicit weighting mechanism to measure its element-wise information quality, which boosts a fine-grained adaptive multimodal fusion. The final multimodal embedding $h_i$ via our proposed multimodal fusion is defined as follows,
\begin{align}
    \begin{split}
        & h^i =  \sum_{m=1}^M  \omega^i_m \odot \mu^i_m,\; \\
        \mathrm{where} \; \omega^i_m = & \;\delta_m^i \cdot \left(\frac{\sigma^i}{\sigma^i_m}\right)^2, \; \frac{1}{\left(\sigma^i\right)^2}= \sum_{m=1}^M \frac{\delta_m^i}{\left(\sigma_m^i\right)^2}.
        \label{eq2}
    \end{split}
\end{align}
$h_i$ is the weighted average of the mean vector $\mu_m$ of the available modalities indicated by $\delta_m$. The weight of $\mu_m$ is determined by its variance $\sigma_m$, where the smaller the variance is, the larger the weight is. 
Through the fine-grained multimodal fusion, the useful information including both complementary and redundant parts can be sufficiently leveraged, which is more robust to the noise corruption and missing modalities. Finally, after fusion, we adopt the shared classifier $f(\cdot)$ on top of $h^i$ to construct the prediction in the following. 
\begin{align}
    \mathcal{L}^M = \mathcal{\ell}\left(y^i,~f\left(h^i\right)\right).
\end{align}
Note that, the reason we use the shared classifier $f(\cdot)$ is that the multimodal representation does share the same latent space with the unimodal representations, as it is a weighted sum of the unimodal representation. Another merit is that we can guarantee the semantic coherence for predictions of each unimodality and multimodality, which further boosts spurious information elimination of the unimodal representation by the implicit supervision from the multimodal representation.

\subsubsection{\textbf{Overall Loss}}
\label{sec3.2.3}
The overall loss function is formulated as
\begin{align}\label{eq:all}
        \mathcal{L} = \mathcal{L}^M + \lambda_1 \mathcal{L}^U + \lambda_2 \mathcal{L}^D,
\end{align}
where $\lambda_1$ and $ \lambda_2$ are hyperparameters used to balance different losses. $\lambda_1$ is to control the unimodal predictive task, usually insensitive within a certain range. $\lambda_2$ that controls the sparsity of embeddings should be properly set, as too large $\lambda_2$ will hurt the unimodal representation learning. Note that, after optimizing the loss $\mathcal{L}$, we directly use $f(h^i)$ only for evaluation. The training procedure of RAML is summarized in Algorithm~\ref{alg:RAML}.

\begin{algorithm}[t!]
    \caption{Training procedure of RAML}
    \label{alg:RAML}
    \begin{algorithmic}
        \STATE {\bfseries Input:} Training data $X$, target set $Y$, missing modality indicator $\delta$, hyperparameters $\lambda_1, \lambda_2$ and batch size $n$.

        \FOR{a sampled batch $ \mathrm{x}= \{x^i\}_{i=1}^n, \, \mathrm{y}= \{y^i\}_{i=1}^n$}
        \FOR{{\bfseries all} $m \in \{1,\cdots,M\}$}
        \STATE $\mu_m=h^{\mu}_m (f_m(\mathrm{x}_m)), \, \sigma_m=h^{\sigma}_m (f_m(\mathrm{x}_m))$ 
        \STATE $q(h_m|\mathrm{x}_m) \sim \mathcal{N}(\mu_m,\sigma_m)$
        \STATE $\mathrm{\hat{y}}_m = f(h_m)\big|_{h_m\sim q(h_m)}$ 
        \ENDFOR
        \STATE $\mathcal{L}^U = \sum_{m=1}^M\mathcal{\ell}(\mathrm{y},\mathrm{\hat{y}}_m)$
        \STATE  $h =  \sum_{m=1}^M  \omega_m \odot \mu_m,\;
        \mathrm{where} \; \omega_m = \delta_m \cdot \left(\frac{\sigma}{\sigma_m}\right)^2, \; \frac{1}{\left(\sigma\right)^2}= \sum_{m=1}^M \frac{\delta_m}{\left(\sigma_m\right)^2}.$ 
        \STATE $\mathrm{\hat{y}} = f(h), \, \mathcal{L}^M = \mathcal{\ell}(\mathrm{y},\mathrm{\hat{y}})$
        \STATE  $\mathcal{L}^{D} = \sum_{m=1}^M \big|\big|(\frac{\sigma}{\sigma_m})^2 \odot \mu_m\big|\big|_1, \,
        \mathrm{where} \; \frac{1}{(\sigma)^2}= \sum_{m=1}^M \frac{1}{(\sigma_m)^2}$
        \STATE // Unimodal embeddings denoising.
        \STATE Optimizing $\mathcal{L} = \mathcal{L}^M + \lambda_1 \mathcal{L}^U + \lambda_2 \mathcal{L}^D$
        
        \ENDFOR
    \end{algorithmic}
\end{algorithm}

\subsection{Discussion}
\subsubsection{\textbf{On the Maximal Predictive Power of RAML}}
In the following, we use an example of a binary classification with two modalities to characterize the maximal predictive power of RAML. 
Specifically, for a certain sample $x^i=\{x^i_1, x^i_2\}$, denote the predictions under the uniform weighting $P_u$, the coarse-grained non-uniform weighting $P_c$, and the fine-grained non-uniform weighting $P_f$ (i.e., the weighting in RAML) for the bimodal representation $\mu^i=\{\mu^i_1, \mu^i_2\}$ respectively as follows,
\begin{align}
   & P_u  = \frac{1}{1  +  \mathrm{exp}\left(- \left(\frac{1}{2} \sum_d \theta^T_d \mu_{1,d}^i + \frac{1}{2} \sum_d \theta^T_d\mu_{2,d}^i\right) \right)}, \nonumber\\
   & P_c  = \frac{1}{1  +  \mathrm{exp}\left(- \left(\alpha^i_1 \sum_d \theta^T_d \mu_{1,d}^i + \alpha^i_2 \sum_d \theta^T_d\mu_{2,d}^i\right) \right)}, \label{eq:power} \\ 
   & P_f = \frac{1}{1  +  \mathrm{exp}\left(- \left(\sum_d \omega^i_{1,d}\theta^T_d \mu_{1,d}^i + \sum_d \omega^i_{2,d} \theta^T_d\mu_{2,d}^i\right) \right)}, \nonumber 
\end{align}
where $\theta = [\theta_1,\cdots,\theta_D] \in \mathbb{R}^D$ is the parameter of the shared predictor $f(\cdot)$, $\alpha_1^i,~\alpha_2^i$ are the coarse-grained weights, and $\omega_{1,d}^i,~\omega_{2,d}^i$ are the fine-grained weights that can exactly reflect the quality of each dimension in the unimodal representation. Then, we have the following theorem that can reflect the maximal predictive power of RAML under the proper conditions.
\begin{theorem}
 Assume that the representation of two modalities in Eq.\eqref{eq:power} have sufficiently captured all the information for prediction. Then, for any predictor $f(\cdot)$ that can properly recognize the pattern of representation $\mu$, if the non-uniform weights are optimized towards the oracle label, we will always have the supremum relation $\sup_\omega P_f\geq \sup_\alpha P_c \geq \sup P_u$, proving that RAML has a better maximal predictive power. 
\end{theorem}

\subsubsection{\textbf{On the Merits and Generality of RAML}}
In terms of the information capture, RAML encourages the sufficient information extraction of each modality by employing a discriminative task with the sparsity constraint, which avoids losing the important redundant and complementary information. From the perspective of multimodal fusion, RAML leverages the element-wise weight to measure the fine-grained quality of the unimodal representations, which promotes the in-depth aggregation of clues for prediction under different corruption conditions. 
In terms of data complexity, RAML only introduces additional $M$ head modules to estimate the variance of the unimodal representation, which is lightweight and efficient in contrast to the consistency-based methods requiring an additional encoder-decoder architecture and the uncertainty-based methods requiring an individual uncertainty estimation network. Besides, since RAML only slightly changes the structure of multimodal models, it is general to many multimodal scenarios and compatible with many existing multimodal fusion methods.

\section{Experiments}
\label{exp}
\subsection{Experimental Setup}
\noindent{\textbf{Datasets.}}
We implement experiments on four multimodal datasets, Food-101~\cite{bossard2014food}, N24News~\cite{wang2021n15news}, IEMOCAP~\cite{busso2008iemocap} and CMU-MOSEI~\cite{zadeh2018multimodal}. 
\begin{itemize}
\item\textbf{Food-101}~\cite{bossard2014food} is a bimodal dataset of 101 food categories. There are 1000 image-text pairs for each category, 750 of which are for training and 250 of which are for testing under human review.

\item\textbf{N24News}~\cite{wang2021n15news} is a bimodal dataset for news classification. It has 61,218 samples of 24 categories, containing both text and image information in each news. We randomly split datasets into training and testing sets following the ratios of 80\% and 20\%. 

\item\textbf{IEMOCAP}~\cite{busso2008iemocap} is a popular multimodal dataset including modalities of audio, text, and video. We follow~\cite{liang2020semi,zhao2021missing} to build the four-class emotion recognition that contains 5531 videos from 10 speakers. A 10-fold cross-validation is adopted, and in each fold, we take the utterances of 8 speakers, 1 speaker, and 1 speaker respectively for training, validation, and testing sets.         

\item\textbf{CMU-MOSEI}~\cite{zadeh2018multimodal} is a large-scale three-modality dataset for sentiment and emotion analysis. This dataset contains 23,453 segments from YouTube with a diverse range of topics. Each segment is composed of audio, text, and video, and is annotated with negative, neutral and positive labels. In CMU-MOSEI, 16,326 samples are used as the training set, and the remaining 1,871 and 4,659 samples are used as validation and testing sets respectively.

\end{itemize}


\noindent{\textbf{Preprocessing.}} All experiments are performed on the features extracted from different pre-trained models.
For the two bimodal datasets Food-101 and N24News, we extract a 512-dimensional vector for each image from the penultimate layer of a pre-trained ResNet-18~\cite{he2016deep} and extract 1024-dimensional vectors for the text modality via a pre-trained BERT-large model~\cite{DBLP:conf/naacl/DevlinCLT19}. 
For the IEMOCAP dataset, we follow \cite{zhao2021missing} to preprocess the data for a fair comparison. In terms of the audio modality, we extract the 130-dimensional frame-level acoustic features by the OpenSMILE toolkit~\cite{eyben2010opensmile} with the configuration of ``IS13 ComPar". In terms of the video modality, the frame-level contextual features are extracted by the BERT-large model. For the video, we extract the facial vectors of 342 dimensions using a pre-trained DenseNet~\cite{huang2017densely}.
For the CMU-MOSEI dataset, we follow \cite{yu2021learning,li2023decoupled} to preprocess the data under the aligned setting. A BERT-base model is exploited to obtain 768-dimensional frame-level contextual features. The video modality is encoded via Facet~\cite{baltruvsaitis2016openface} to represent the presence of the total 35 facial units. The acoustic modality is processed by COVAREP~\cite{degottex2014covarep} to obtain the 74-dimensional frame-level features.

\definecolor{greyL}{RGB}{235,235,235}
\begin{table*}[t!]
\centering
\caption{Performance under different missing-modality conditions, where ``I", ``T", ``A", and ``V" respectively represent the image, text, audio, and video modality available during testing. “Average” refers to the average performance over all the possible conditions. Following~\cite{zhao2021missing}, on Food-101, N24News and CMU-MOSEI, the metric is WA, and on IEMOCAP, it is UA. The best results are in bold and the second-best ones are marked with underline.}
\label{tab:1}
\resizebox{0.97\textwidth}{!}{
\setlength{\tabcolsep}{3.0mm}{
\begin{tabular}{c|cccccccc}
\toprule[1.2pt]    
                      & \multicolumn{4}{c|}{Food-101}                                                              & \multicolumn{4}{c}{N24News}                                            \\
Method                & \{I\}           & \{T\}           & \{I,T\}         & \multicolumn{1}{c|}{Average}         & \{I\}           & \{T\}           & \{I,T\}         & Average          \\ 
\midrule[0.2pt]
Concat Fusion         & 0.4115          & 0.3298          & 0.7060          & \multicolumn{1}{c|}{0.4824}          & 0.3973          & 0.2811          & 0.5228          & 0.4004           \\
Augmentation          & 0.4155          & 0.3310          & 0.7076          & \multicolumn{1}{c|}{0.4847}          & 0.3893          & 0.3377          & 0.5477          & 0.4249           \\ 
\midrule[0.2pt]
MVAE                  & 0.4759          & 0.4784          & 0.7226          & \multicolumn{1}{c|}{0.5589}          & 0.4048          & 0.3655          & 0.5451          & 0.4385           \\
MMVAE                 & 0.4989          & 0.4719          & 0.7130          & \multicolumn{1}{c|}{0.5613}          & 0.4051          & 0.3710          & 0.5391          & 0.4379           \\ 
\midrule[0.2pt]
CRA                   & 0.4482          & 0.5396          & 0.7188          & \multicolumn{1}{c|}{0.5689}          & 0.3979          & 0.3833          & 0.5493          & 0.4435           \\
MMIN                  & 0.4754          & 0.5417          & 0.7309          & \multicolumn{1}{c|}{0.5826}          & 0.3928          & 0.4068          & 0.5485          & 0.4494           \\
TATE                  & 0.4195          & 0.3844          & 0.7079          & \multicolumn{1}{c|}{0.5039}          & 0.3806          & 0.3320          & 0.5392          & 0.4172           \\ 
\midrule[0.2pt]
TMC                   & 0.4918          & 0.5725          & 0.7203          & \multicolumn{1}{c|}{0.5949}          & 0.4062          & 0.3935          & 0.5206          & 0.4401           \\
MD                    & \underline{0.5057}          & 0.5855          & \underline{0.7604}          & \multicolumn{1}{c|}{\underline{0.6172}}          & 0.3938          & \cellcolor{greyL}\textbf{0.4147} & \underline{0.5499}          & \underline{0.4528}           \\
ETMC                  & 0.4952          & \underline{0.5938}          & \underline{0.7604}          & \multicolumn{1}{c|}{0.6165}          & \underline{0.4016}          & 0.3958          & 0.5381          & 0.4452           \\ 
\midrule[0.2pt]
RAML                  & \cellcolor{greyL}\textbf{0.5109} & \cellcolor{greyL}\textbf{0.6646} & \cellcolor{greyL}\textbf{0.7864} & \multicolumn{1}{c|}{\cellcolor{greyL}\textbf{0.6540}} & \cellcolor{greyL}\textbf{0.4106} & \underline{0.4141}          & \cellcolor{greyL}\textbf{0.5641} & \cellcolor{greyL}\textbf{0.4630}  \\ 
$\Delta$                  &\textcolor{mydarkred}{\;0.52\% $\!\!\uparrow$} & \textcolor{mydarkred}{\;7.91\% $\!\!\uparrow$}          & \textcolor{mydarkred}{\;2.60\% $\!\!\uparrow$} & \multicolumn{1}{c|}{\textcolor{mydarkred}{\;3.68\% $\!\!\uparrow$}}                      & \textcolor{mydarkred}{\;0.90\% $\!\!\uparrow$} & \textcolor{mydarkblue}{\;0.06\% $\!\!\downarrow$} & \textcolor{mydarkred}{\;1.42\% $\!\!\uparrow$} & \textcolor{mydarkred}{\;1.02\% $\!\!\uparrow$}   \\
\midrule[0.6pt]\midrule[0.6pt]
                      & \multicolumn{8}{c}{IEMOCAP}                                         \\
Method                & \{A\}           & \{T\}           & \{V\}           & \{A,T\}                              & \{A,V\}         & \{T,V\}         & \{A,T,V\}       & Average          \\ 
\midrule[0.2pt]
Concat Fusion         & 0.4124          & 0.5843          & 0.4119          & 0.7201                               & 0.5692          & 0.6248          & 0.7788          & 0.5859           \\
Augmentation          & 0.4070          & 0.6016          & 0.4182          & 0.7361                               & 0.5518          & 0.6351          & 0.7809          & 0.5901           \\ 
\midrule[0.2pt]
MVAE                  & 0.5637          & 0.6835          & 0.4999          & 0.7233                               & 0.6499          & 0.7062          & 0.7342          & 0.6515           \\
MMVAE                 & 0.5691          & 0.6794          & 0.5131          & 0.7461                               & 0.6553          & 0.7242          & 0.7810          & 0.6669           \\ 
\midrule[0.2pt]
CRA                   & 0.5693          & 0.6819          & 0.4902          & 0.7504                               & \underline{0.6559}          & 0.7298          & 0.7760          & 0.6648           \\
MMIN                  & 0.5900          & 0.6802          & \underline{0.5160}          & 0.7514                               & 0.6543          & 0.7361          & \underline{0.7812}          & 0.6727           \\
TATE                  & 0.5564          & 0.6630          & 0.5149          & 0.7327                               & 0.6471          & 0.7147          & 0.7570          & 0.6508           \\ 
\midrule[0.2pt]
TMC                   & 0.5904          & \underline{0.6945}          & 0.4841          & 0.7460                               & 0.6268          & 0.7230          & 0.7669          & 0.6617           \\
MD                    & 0.5910          & 0.6926          & 0.5121          & \underline{0.7561}                               & 0.6459          & \underline{0.7363}          & 0.7788          & \underline{0.6732}           \\
ETMC                  & \underline{0.5919}          & 0.6893          & 0.5076          & 0.7502                               & 0.6543          & 0.7312          & 0.7789          & 0.6719           \\ 
\midrule[0.2pt]
RAML                  & \cellcolor{greyL}\textbf{0.6060} & \cellcolor{greyL}\textbf{0.6952}          & \cellcolor{greyL}\textbf{0.5274} & \cellcolor{greyL}\textbf{0.7572}                      & \cellcolor{greyL}\textbf{0.6710} & \cellcolor{greyL}\textbf{0.7480} & \cellcolor{greyL}\textbf{0.7868} & \cellcolor{greyL}\textbf{0.6845}  \\
$\Delta$                  &\textcolor{mydarkred}{\;1.41\% $\!\!\uparrow$} & \textcolor{mydarkred}{\;0.07\% $\!\!\uparrow$}          & \textcolor{mydarkred}{\;1.14\% $\!\!\uparrow$} & \textcolor{mydarkred}{\;0.11\% $\!\!\uparrow$}                      & \textcolor{mydarkred}{\;1.51\% $\!\!\uparrow$} & \textcolor{mydarkred}{\;1.17\% $\!\!\uparrow$} & \textcolor{mydarkred}{\;0.56\% $\!\!\uparrow$} & \textcolor{mydarkred}{\;1.13\% $\!\!\uparrow$}   \\
\midrule[0.6pt]\midrule[0.6pt]
 & \multicolumn{8}{c}{CMU-MOSEI}                                                                                                                                       \\
Method                & \{A\}           & \{T\}           & \{V\}           & \{A,T\}                              & \{A,V\}         & \{T,V\}         & \{A,T,V\}       & Average          \\ 
\midrule[0.2pt]
Concat Fusion         & 0.2206          & \underline{0.6757}          & 0.4881          & 0.6757                               & 0.4370         & 0.6780          & 0.6817          & 0.5510           \\
Augmentation          & 0.2198          & 0.6697          & 0.4703          & 0.6688                               & 0.4213          & 0.6684          & 0.6791          & 0.5425           \\ 
\midrule[0.2pt]
MVAE                  & 0.4902          & 0.6740          & 0.4924          & 0.6718                               & 0.4902          & 0.6727          & 0.6729          & 0.5949           \\
MMVAE                 & 0.4902          & 0.6542          & 0.4902          & 0.6531                               & 0.4898          & 0.6564          & 0.6574          & 0.5845           \\ 
\midrule[0.2pt]
CRA                   & 0.4926          & 0.6707          & 0.4937          & 0.6720                               & 0.5018          & 0.6763          & 0.6765          & 0.5977           \\
MMIN                  & 0.4904          & 0.6705          & 0.4999          & 0.6748                               & 0.5046          & 0.6733          & 0.6783          & 0.5988           \\
TATE                  & 0.4902          & 0.6746          & 0.5044          & 0.6744                               & 0.5055          & 0.6755          & 0.6780          & 0.6004           \\ 
\midrule[0.2pt]
TMC                   & \underline{0.4939}          & 0.6733         & 0.4937          & \underline{0.6780}                               & 0.5008          & \underline{0.6808}          & \underline{0.6843}          & 0.6007           \\
MD                    & 0.4902          & 0.6688          & \cellcolor{greyL}\textbf{0.5063}          & 0.6735                               & 0.5080          & 0.6806          & 0.6808          & 0.6012           \\
ETMC                  & 0.4926          & 0.6720          & \underline{0.5053}          & 0.6722                               & \underline{0.5168}          & 0.6759          & 0.6768          & \underline{0.6017}           \\ 
\midrule[0.2pt]
RAML                  & \cellcolor{greyL}\textbf{0.5083} & \cellcolor{greyL}\textbf{0.6808}          & 0.5029 & \cellcolor{greyL}\textbf{0.6787}                      & \cellcolor{greyL}\textbf{0.5194} & \cellcolor{greyL}\textbf{0.6841} & \cellcolor{greyL}\textbf{0.6879} & \cellcolor{greyL}\textbf{0.6089}  \\ 
$\Delta$                  &\textcolor{mydarkred}{\;1.44\% $\!\!\uparrow$} & \textcolor{mydarkred}{\;0.51\% $\!\!\uparrow$}          & \textcolor{mydarkblue}{\;0.34\% $\!\!\downarrow$} & \textcolor{mydarkred}{\;0.07\% $\!\!\uparrow$}                      & \textcolor{mydarkred}{\;0.26\% $\!\!\uparrow$} & \textcolor{mydarkred}{\;0.33\% $\!\!\uparrow$} & \textcolor{mydarkred}{\;0.36\% $\!\!\uparrow$} & \textcolor{mydarkred}{\;0.72\% $\!\!\uparrow$}   \\ 
\bottomrule[1.2pt]    
\end{tabular}}}
\end{table*}

\definecolor{greyL}{RGB}{235,235,235}
\begin{table*}[t!]
\caption{Fusion Performance comparison when adding Gaussian noise on each image (top half of the table) or masking words on each text (bottom half of the table). The variance $\sigma$ of the Gaussian noise changes from 0.08 to 0.38 and the probability $p$ of masking words increases from 0.05 to 0.25. 
}
\centering
\label{tab:noise}
\resizebox{0.99\textwidth}{!}{
\setlength{\tabcolsep}{0.95mm}{
\begin{tabular}{c|ccccc|ccccc}
\toprule[1.2pt]
                                     & \multicolumn{5}{c|}{Food-101}                                                           & \multicolumn{5}{c}{N24News}                                                             \\ 
 Method        & $\sigma=0.08$   & $\sigma=0.12$   & $\sigma=0.18$   & $\sigma=0.26$   & $\sigma=0.38$   & $\sigma=0.08$   & $\sigma=0.12$   & $\sigma=0.18$   & $\sigma=0.26$   & $\sigma=0.38$   \\ \midrule[0.2pt] 
                                  Concat Fusion & 0.6615          & 0.6300          & 0.5832          & 0.5198          & 0.4402          & 0.5130          & 0.5053          & 0.4853          & 0.4564          & 0.3993          \\ 
                                  Augmentation &  0.6656          & 0.6338          & 0.5865          & 0.5173          & 0.4385          & 0.5403          & 0.5285          & 0.5079          & \underline{0.4844}          & 0.4340          \\ \midrule[0.2pt] 
                                  MVAE          & 0.6958          & 0.6740          & 0.6393          & 0.5940          & 0.5420          & 0.5367          & 0.5258          & \underline{0.5090}          & 0.4836          & \underline{0.4439}          \\
                                  MMVAE         & 0.6814          & 0.6549          & 0.6186          & 0.5689          & 0.5118          & 0.5251          & 0.5168          & 0.4976          & 0.4663          & 0.4207          \\ \midrule[0.2pt] 
                                  CRA           & 0.6947          & 0.6762          & 0.6500          & 0.6129          & 0.5725          & 0.5366          & \underline{0.5258}          & 0.5072          & 0.4757          & 0.4256          \\
                                  MMIN          & 0.7003          & 0.6805          & 0.6478          & 0.6084          & 0.5665          & \underline{0.5372}          & 0.5257          & 0.5086          & 0.4823          & 0.4316          \\
                                  TATE          & 0.6646          & 0.6349          & 0.5914          & 0.5370          & 0.4745          & 0.5275          & 0.5174          & 0.4983          & 0.4684          & 0.4260          \\ \midrule[0.2pt] 
                                  TMC           & 0.6919          & 0.6718          & 0.6433          & 0.6114          & 0.5779          & 0.5089          & 0.5010          & 0.4849          & 0.4606          & 0.4201          \\
                                  MD            & 0.7234          & 0.6928          & 0.6487          & 0.5924          & 0.5196          & 0.5350          & 0.5249          & 0.5030          & 0.4723          & 0.4242          \\
                                  ETMC          & \underline{0.7332}          & \underline{0.7117}          & \underline{0.6810}          & \underline{0.6503}          & \underline{0.6124}          & 0.5267          & 0.5179          & 0.5047          & 0.4722          & 0.4287          \\ \midrule[0.2pt] 
                                  RAML           & \cellcolor{greyL}\textbf{0.7619} & \cellcolor{greyL}\textbf{0.7475} & \cellcolor{greyL}\textbf{0.7260} & \cellcolor{greyL}\textbf{0.6981} & \cellcolor{greyL}\textbf{0.6661} & \cellcolor{greyL}\textbf{0.5525} & \cellcolor{greyL}\textbf{0.5393} & \cellcolor{greyL}\textbf{0.5246} & \cellcolor{greyL}\textbf{0.4984} & \cellcolor{greyL}\textbf{0.4571} \\ 
                                  $\Delta$           & \textcolor{mydarkred}{\;2.87\% $\!\!\uparrow$} & \textcolor{mydarkred}{\;3.58\% $\!\!\uparrow$} & \textcolor{mydarkred}{\;4.50\% $\!\!\uparrow$} & \textcolor{mydarkred}{\;4.78\% $\!\!\uparrow$} & \textcolor{mydarkred}{\;5.37\% $\!\!\uparrow$} & \textcolor{mydarkred}{\;1.53\% $\!\!\uparrow$} & \textcolor{mydarkred}{\;1.35\% $\!\!\uparrow$} & \textcolor{mydarkred}{\;1.56\% $\!\!\uparrow$} & \textcolor{mydarkred}{\;1.40\% $\!\!\uparrow$} & \textcolor{mydarkred}{\;1.32\% $\!\!\uparrow$} \\ 
                                  \midrule[0.6pt]\midrule[0.6pt]
 Method        & $p=0.05$          & $p=0.1$           & $p=0.15$          & $p=0.2$           & $p=0.25$          & $p=0.05$          & $p=0.1$           & $p=0.15$          & $p=0.2$           & $p=0.25$          \\ \midrule[0.2pt] 
                                  Concat Fusion & 0.6913          & 0.6701          & 0.6482          & 0.6239          & 0.5985          & 0.5221          & 0.5171          & 0.5034          & 0.4907          & 0.4748          \\ 
                                  Augmentation & 0.6924          & 0.6728          & 0.6501          & 0.6239          & 0.5997          & 0.5457          & 0.5388          & 0.5300          & 0.5191          & 0.4960         \\ \midrule[0.2pt] 
                                  MVAE          & 0.7069          & 0.6859          & 0.6638          & 0.6383          & 0.6124          & 0.5402          & 0.5379          & 0.5219          & 0.5210          & 0.4911          \\
                                  MMVAE         & 0.7000          & 0.6818          & 0.6605          & 0.6403          & 0.6149          & 0.5330          & 0.5280          & 0.5208          & 0.5057          & 0.4951          \\ \midrule[0.2pt] 
                                  CRA           & 0.6853          & 0.6481          & 0.6089          & 0.5740          & 0.5335          & \underline{0.5479}          & 0.5362          & 0.5286          & 0.5156          & 0.5026          \\
                                  MMIN          & 0.7027          & 0.6700          & 0.6370          & 0.6010          & 0.5640          & 0.5459          & \underline{0.5402}          & 0.5259          & 0.5138          & 0.5002          \\ \midrule[0.2pt] 
                                  TATE          & 0.6829          & 0.6605          & 0.6282          & 0.5987          & 0.5663          & 0.5366          & 0.5239          & 0.5153          & 0.5031          & 0.4821          \\
                                  TMC           & 0.7021          & 0.6796          & 0.6541          & 0.6314          & 0.6038          & 0.5190          & 0.5138          & 0.5028          & 0.4918          & 0.4819          \\ \midrule[0.2pt] 
                                  MD            & 0.7396          & 0.7155          & 0.6941          & 0.6707          & 0.6414          & 0.5465          & 0.5399          & \underline{0.5309}          & \underline{0.5216}          & 0.5051          \\
                                  ETMC          & \underline{0.7440}          & \underline{0.7216}          & \underline{0.6982}          & \underline{0.6756}          & \underline{0.6468}          & 0.5143          & 0.5110          & 0.4991          & 0.4898          & 0.4792          \\ \midrule[0.2pt] 
                                  RAML           & \cellcolor{greyL}\textbf{0.7655} & \cellcolor{greyL}\textbf{0.7443} & \cellcolor{greyL}\textbf{0.7171} & \cellcolor{greyL}\textbf{0.6859} & \cellcolor{greyL}\textbf{0.6549} & \cellcolor{greyL}\textbf{0.5575}          & \cellcolor{greyL}\textbf{0.5533}          & \cellcolor{greyL}\textbf{0.5382}          & \cellcolor{greyL}\textbf{0.5264}          & \cellcolor{greyL}\textbf{0.5113}           \\ 
                                  $\Delta$           & \textcolor{mydarkred}{\;2.15\% $\!\!\uparrow$} & \textcolor{mydarkred}{\;2.27\% $\!\!\uparrow$} & \textcolor{mydarkred}{\;1.89\% $\!\!\uparrow$} & \textcolor{mydarkred}{\;1.03\% $\!\!\uparrow$} & \textcolor{mydarkred}{\;0.81\% $\!\!\uparrow$} & \textcolor{mydarkred}{\;0.96\% $\!\!\uparrow$} & \textcolor{mydarkred}{\;1.31\% $\!\!\uparrow$} & \textcolor{mydarkred}{\;0.73\% $\!\!\uparrow$} & \textcolor{mydarkred}{\;0.48\% $\!\!\uparrow$} & \textcolor{mydarkred}{\;0.62\% $\!\!\uparrow$} \\ 
                                  \bottomrule[1.2pt] 
\end{tabular}}}
\end{table*}

\noindent{\textbf{Modality Encoder.}} 
The modality encoders on top of the pre-extracted features are clarified as follows.
We apply two two-layer Multi-Layer Perceptron (MLP) networks to capture the image and text information respectively.
For sequential features in the IEMOCAP datasets, we follow~\cite{zhao2021missing} to employ a TextCNN~\cite{DBLP:conf/emnlp/Kim14} to acquire the utterance-level textual embeddings and adopt two one-layer LSTM~\cite{sak2014long} networks to capture the temporal acoustic and facial information based on the corresponding frame-level raw features respectively. Then, the max-pooling operation is used to compute the utterance-level acoustic and facial embeddings. For the CMU-MOSEI datasets, three one-layer LSTM networks with the max-pooling operation are used to encode the utterance-level textual, acoustic and facial embeddings respectively.

\noindent{\textbf{Experimental Details. }}For all experiments, we use the Adam optimizer~\cite{kingma2014adam} and set the batch size as 128. The dimension of the Gaussian distribution is 128 and we set the learning rate to 1e-3 with a weight decay of 5e-4. 
During training, we augment full-modality samples with all possible missing patterns to simulate the missing modality conditions. For bimodal datasets, there are three cases: missing image, missing text, and missing none. For trimodal datasets, there are seven missing cases.
Following~\cite{zhao2021missing,DBLP:conf/cvpr/0002R0T022}, on Food-101, N24News and CMU-MOSEI, we use the weighted accuracy evaluation metrics (WA), and on IEMOCAP unweighted accuracy (UA) is applied to measure the performance on the test set.

\subsection{Performance under Imperfect Modalities} \label{sec:imperfect}
To evaluate the robustness of the proposed method, we choose the following baselines in the comparison: 
1) Concat Fusion: A simple intermediate fusion model which concatenates embeddings from different modalities;
2) Augmentation: a data augmentation-based method which augments missing-modality data based on Concat Fusion;
3) Uncertainty-based methods TMC~\cite{DBLP:conf/iclr/HanZFZ21}, MD~\cite{han2022multimodal}, ETMC~\cite{DBLP:journals/pami/HanZFZ23}; 
4) Consistency-based methods MVAE~\cite{wu2018multimodal} and MMVAE~\cite{shi2019variational} from the perspective of reconstruction, and CRA~\cite{tran2017missing}, MMIN~\cite{zhao2021missing}, TATE~\cite{zeng2022tag} from the joint representation learning.

\subsubsection{\textbf{Robustness to Missing Modality}} We construct different testing subsets corresponding to different missing-modality conditions respectively on all testing samples to evaluate the robustness.
Table~\ref{tab:1} presents the experimental results of RAML and other SOTA methods under different missing-modality testing conditions. We can see that the `Concat Fusion' method only trained on full-modality samples suffers from a significant performance drop when encountering missing modality testing conditions. For example, the accuracy decreases by 37.62\% on Food-101 when the image modality is missing and by 19.45\% on IEMOCAP when only the text modality is available. This indicates that the traditional multimodal fusion method learned under the full-modality condition is very sensitive to the missing modality problem. Another interesting finding is that the `Augmentation' method is not superior to `Concat Fusion' in CMU-MOSEI, which means that augmenting the simulated missing-modality data only during training does not always guarantee the improvement of robustness. This could be attributed to the complex and noisy data of some modalities in CMU-MOSEI.
Comparing the results of different consistency-based and uncertainty-based methods, RAML achieves the best results in almost all the settings on four multimodal datasets. 
The performance of RAML improves by 3.68\%, 1.02\%, 1.13\% and 0.72\% on average compared to the second-best method on Food-101, N24News, IEMOCAP and CMU-MOSEI respectively. These results illustrate the effectiveness of our proposed method in adaptively perceiving and aggregating the information from different multimodal combinations.

\subsubsection{\textbf{Robustness to Singly-Corrupted Modality}}
In order to verify the robustness under one corrupted input, we conduct experiments on the testing data with the corrupted image or text modality on Food-101 and N24News. We corrupt the image modality by adding Gaussian noise on each image under the five noise levels following~\cite{michaelis2019dragon}. And the corruption of the text modality is simulated by masking partial words within a certain probability in each text.   
The masking probability increases from 0.05 to 0.25, increasing by 0.05 each time. We summarize the results in Table~\ref{tab:noise}, and as can be seen, RAML achieves the best results in all noise environments on the two datasets. For example, when the variance of Gaussian noise is 0.18, our method is 4.50\% higher and 1.56\% higher than the second-best method on Food-101 and N24News respectively. When the masking probability is 0.1, our method outperforms the second-best method by 2.27\% on Food-101 and 1.31\% on N24News than the second-best method. All these results indicate that our method can effectively handle the varying input noise by perceiving the informative and uninformative contents and assigning weights to these contents adaptively.

\begin{table}[t!]
\centering
\caption{Fusion performance comparison under Gaussian noise on images and masking words on text simultaneously. At the top half of the table, the $\sigma$ of Gaussian noise changes from 0.08 to 0.38 while the probability of masking words is fixed as 0.1. At the bottom half of the table, the probability of masking words varies from 0.05 to 0.25 while the $\sigma$ of Gaussian noise changes is fixed as 0.18.}
\label{tab:noise*2}
\resizebox{0.99\linewidth}{!}{
\setlength{\tabcolsep}{0.2mm}{
\begin{tabular}{c|ccccc}
\toprule[1.2pt]
Method        & $\sigma=0.08$   & $\sigma=0.12$   & $\sigma=0.18$   & $\sigma=0.26$   & $\sigma=0.38$   \\ \midrule[0.2pt]
Concat Fusion & 0.6222          & 0.5876          & 0.5352          & 0.4705          & 0.3938          \\ 
Augmentation  & 0.6258          & 0.5921          & 0.5381          & 0.4698          & 0.3898          \\\midrule[0.2pt]
MVAE          & 0.6561          & 0.6323          & 0.5929          & 0.5414          & 0.4827          \\
MMVAE         & 0.6461          & 0.6180          & 0.5744          & 0.5210          & 0.4593          \\ \midrule[0.2pt]
CRA           & 0.6189          & 0.5980          & 0.5654          & 0.5258          & 0.4802          \\
MMIN          & 0.6298          & 0.6062          & 0.5689          & 0.5261          & 0.4801          \\
TATE          & 0.6115          & 0.5785          & 0.5303          & 0.4730          & 0.4064          \\ \midrule[0.2pt]
TMC           & 0.6446          & 0.6225          & 0.5888          & 0.5506          & 0.5136          \\
MD            & 0.6683          & 0.6336          & 0.5836          & 0.5239          & 0.4481          \\
ETMC          & \underline{0.6890}          & \underline{0.6639}          & \underline{0.6267}          & \underline{0.5896}          & \underline{0.5478}          \\ \midrule[0.2pt]
RAML           & \cellcolor{greyL}\textbf{0.7135} & \cellcolor{greyL}\textbf{0.6943} & \cellcolor{greyL}\textbf{0.6653} & \cellcolor{greyL}\textbf{0.6314} & \cellcolor{greyL}\textbf{0.5967} \\
$\Delta$       & \textcolor{mydarkred}{\;2.45\% $\!\!\uparrow$} & \textcolor{mydarkred}{\;3.04\% $\!\!\uparrow$} & \textcolor{mydarkred}{\;3.86\% $\!\!\uparrow$} & \textcolor{mydarkred}{\;4.18\% $\!\!\uparrow$} & \textcolor{mydarkred}{\;4.89\% $\!\!\uparrow$} \\ 
\midrule[0.6pt]\midrule[0.6pt]
Method        & $p=0.05$          & $p=0.1$           & $p=0.15$          & $p=0.2$           & $p=0.25$   \\ \midrule[0.2pt]
Concat Fusion & 0.6108          & 0.5876          & 0.5626         & 0.5364          & 0.5074          \\ 
Augmentation  & 0.6131          & 0.5921          & 0.5651          & 0.5399          & 0.5092          \\\midrule[0.2pt]
MVAE          & 0.6524          & 0.6323          & 0.6035          & 0.5773          & 0.5455          \\
MMVAE         & 0.6382          & 0.6180          & 0.5964          & 0.5673          & 0.5384          \\ \midrule[0.2pt]
CRA           & 0.6389          & 0.5980          & 0.5553          & 0.5180          & 0.4772          \\
MMIN          & 0.6461          & 0.6062          & 0.5701          & 0.5309          & 0.4836          \\
TATE          & 0.6067          & 0.5785          & 0.5450          & 0.5136          & 0.4769          \\ \midrule[0.2pt]
TMC           & 0.6503          & 0.6225          & 0.5908          & 0.5626          & 0.5313          \\
MD            & 0.6669          & 0.6336          & 0.6086          & 0.5785          & 0.5410          \\
ETMC          & \underline{0.6904}          & \underline{0.6639}          & \underline{0.6346}          & \underline{0.6034}          & \underline{0.5722}          \\ \midrule[0.2pt]
RAML           & \cellcolor{greyL}\textbf{0.7226} & \cellcolor{greyL}\textbf{0.6943} & \cellcolor{greyL}\textbf{0.6616} & \cellcolor{greyL}\textbf{0.6262} & \cellcolor{greyL}\textbf{0.5825} \\
$\Delta$       & \textcolor{mydarkred}{\;3.22\% $\!\!\uparrow$} & \textcolor{mydarkred}{\;3.04\% $\!\!\uparrow$} & \textcolor{mydarkred}{\;2.70\% $\!\!\uparrow$} & \textcolor{mydarkred}{\;2.28\% $\!\!\uparrow$} & \textcolor{mydarkred}{\;2.03\% $\!\!\uparrow$} \\ 

\bottomrule[1.2pt] 
\end{tabular}}}
\end{table}

\subsubsection{\textbf{Robustness to Dual-Corrupted Modalities}} In addition to assessing the robustness of the model when one modality is corrupted, we conduct the experiment to evaluate its robustness when both modalities are corrupted on Food-101. To be specific, we maintain the masking probability of 0.1 for the text modality and vary the variance of Gaussian noise on the image modality from 0.08 to 0.38. Alternatively, we perform experiments with a fixed Gaussian noise variance of 0.18, adjusting the masking probability within a range of 0.05 to 0.25. The results of this experiment are presented in Table~\ref{tab:noise*2}. RAML achieves superior accuracy under all noise conditions, even when the noise is serious. For example, with a Gaussian noise variance of 0.38 and a masking probability of 0.1, our method demonstrates a 4.89\% improvement over the second-best method. Similarly, with a Gaussian noise variance of 0.18 and a masking probability of 0.25, our method outperforms the second-best method by 2.03\%.
All these findings suggest that the model is capable of maintaining robust performance even when both modalities are subject to corruption.

\subsection{Performance under Full Modalities}\label{sec:full}
To show the promise of our method under full modalities, we conduct experiments to compare RAML with some representative baselines, namely, Concat Fusion and the uncertainty-based methods TMC, MD and ETMC in the full-modality condition. Notice that different from the experiments in Section~\ref{sec:imperfect}, in this part, only full-modality samples are used during both training and testing.

According to the results in Table~\ref{tab:full}, we can observe that RAML achieves the best performance on Food-101 and N24News, and competes well on IEMOCAP and CMU-MOSEI. Specifically, RAML outperforms the second-best method by 1.58\% on Food-101, 0.95\% on N24News, 0.32\% on IEMOCAP, 0.15\% on CMU-MOSEI. This indicates that RAML can not only deal with various cases of data imperfection, but also has the comparable or even better ability of multimodal information aggregation as many existing fusion methods. We suggest that this information aggregation ability may result from lossless learning of unimodal information.

\begin{table}[!t]
\caption{Comparison of RAML and uncertainty-based methods under full-modality training.}
\centering
\label{tab:full}
\resizebox{0.99\linewidth}{!}{
\setlength{\tabcolsep}{0.9mm}{
\begin{tabular}{c|cccc}
\toprule[1.2pt]
Method & Food-101 & N24News & IEMOCAP & CMU-MOSEI \\ \midrule[0.2pt]
Concat Fusion & 0.7060 & 0.5228 & 0.7788 & \underline{0.6817} \\
TMC    & 0.7242   & 0.5190  & 0.7727 & 0.6783 \\
MD     & 0.7412   & \underline{0.5500}  & 0.7974 & 0.6813\\
ETMC   & \underline{0.7655 }  & 0.5393  & \underline{0.7986} & 0.6806\\
RAML    & \cellcolor{greyL}\textbf{0.7813}   & \cellcolor{greyL}\textbf{0.5595}  & \cellcolor{greyL}\textbf{0.8018} & \cellcolor{greyL}\textbf{0.6832}\\ \bottomrule[1.2pt] 
\end{tabular}}}
\end{table}

\begin{table}[t!]
\caption{Different weighting strategies among modalities.
}
\centering
\label{tab:weight}
\resizebox{0.99\linewidth}{!}{
\setlength{\tabcolsep}{0.4mm}{
\begin{tabular}{c|cccc}
\toprule[1.2pt]
               & Food-101        & N24News         & IEMOCAP   & CMU-MOSEI\\ \midrule[0.2pt]
Identical        & 0.6346          & 0.4558          & 0.6639  & 0.5935  \\
Fixed          & 0.6369          & 0.4590          & 0.6722    & 0.6024 \\
Adaptive (RAML) & \cellcolor{greyL}\textbf{0.6540} & \cellcolor{greyL}\textbf{0.4630} & \cellcolor{greyL}\textbf{0.6845} & \cellcolor{greyL}\textbf{0.6089} \\ \bottomrule[1.2pt] 
\end{tabular}}}
\end{table}


\begin{table*}[!t]
\caption{Ablation study on the Food-101, N24News, and IEMOCAP datasets. For clarity, $\times$ and $\checkmark$ in the table indicate without and with the corresponding loss term respectively.
}
\centering
\label{tab:3}
\resizebox{0.92\textwidth}{!}{
\setlength{\tabcolsep}{2.7mm}{
\begin{tabular}{ccc|cccccccc}
\toprule[1.2pt]
\textbf{}       & \textbf{}       & \textbf{}       & \multicolumn{4}{c|}{Food-101}                                                             & \multicolumn{4}{c}{N24News}                                            \\
$\mathcal{L}^M$ & $\mathcal{L}^U$ & $\mathcal{L}^D$ & \{I\}           & \{T\}           & \{I,T\}         & \multicolumn{1}{c|}{Average}                             & \{I\}           & \{T\}           & \{I,T\}         & Average          \\ 
\midrule[0.2pt]
$\checkmark$    & $\times$        & $\times$        & 0.4915          & 0.4879          & 0.7175          & \multicolumn{1}{c|}{0.5656}                              & 0.4066          & 0.3736          & 0.5441          & 0.4414           \\
$\checkmark$    & $\times$        & $\checkmark$    & 0.4928          & 0.5048          & 0.7301          & \multicolumn{1}{c|}{0.5759}                              & 0.3987          & 0.3864          & 0.5527          & 0.4459           \\
$\checkmark$    & $\checkmark$    & $\times$        & \cellcolor{greyL}\textbf{0.5165} & 0.6147          & 0.7533          & \multicolumn{1}{c|}{0.6281}                              & 0.4063          & 0.4090          & 0.5617          & 0.4590           \\
$\checkmark$    & $\checkmark$    & $\checkmark$    & 0.5109          & \cellcolor{greyL}\textbf{0.6646} & \cellcolor{greyL}\textbf{0.7864} & \multicolumn{1}{c|}{\cellcolor{greyL}\textbf{0.6540}}                     & \cellcolor{greyL}\textbf{0.4106} & \cellcolor{greyL}\textbf{0.4141} & \cellcolor{greyL}\textbf{0.5641} & \cellcolor{greyL}\textbf{0.4630}  \\ 
\midrule[0.6pt]\midrule[0.6pt]
                &                 &                 & \multicolumn{8}{c}{IEMOCAP}                                                                                                                                        \\
$\mathcal{L}^M$ & $\mathcal{L}^U$ & $\mathcal{L}^D$ & \{A\}           & \{T\}           & \{V\}           & \multicolumn{1}{c}{\{A,T\}}         & \{A,V\}         & \{T,V\}         & \{A,T,V\}       & Average          \\ 
\midrule[0.2pt]
$\checkmark$    & $\times$        & $\times$        & 0.5013          & 0.5991          & 0.4375          & \multicolumn{1}{c}{0.7469}          & 0.6115          & 0.6360          & 0.7919          & 0.6177           \\
$\checkmark$    & $\times$        & $\checkmark$    & 0.5403          & 0.6613          & 0.4662          & \multicolumn{1}{c}{0.7579}          & 0.6375          & 0.7106          & 0.7928          & 0.6524           \\
$\checkmark$    & $\checkmark$    & $\times$        & 0.5859          & 0.6897          & 0.5108          & \multicolumn{1}{c}{\cellcolor{greyL}\textbf{0.7628}} & 0.6684          & 0.7249          & \cellcolor{greyL}\textbf{0.7934} & 0.6766           \\
$\checkmark$    & $\checkmark$    & $\checkmark$    & \cellcolor{greyL}\textbf{0.6060} & \cellcolor{greyL}\textbf{0.6952} & \cellcolor{greyL}\textbf{0.5274} & \multicolumn{1}{c}{0.7572}          & \cellcolor{greyL}\textbf{0.6710} & \cellcolor{greyL}\textbf{0.7480} & 0.7868          & \cellcolor{greyL}\textbf{0.6845}  \\ \midrule[0.6pt]\midrule[0.6pt]
                &                 &                 & \multicolumn{8}{c}{CMU-MOSEI}                                                                                                                                        \\
$\mathcal{L}^M$ & $\mathcal{L}^U$ & $\mathcal{L}^D$ & \{A\}           & \{T\}           & \{V\}           & \multicolumn{1}{c}{\{A,T\}}         & \{A,V\}         & \{T,V\}         & \{A,T,V\}       & Average          \\ 
\midrule[0.2pt]
$\checkmark$    & $\times$        & $\times$        & 0.4909          & 0.6667          & 0.4920          & \multicolumn{1}{c}{0.6703}          & 0.4915          & 0.6682          & 0.6637         & 0.5919           \\
$\checkmark$    & $\times$        & $\checkmark$    & 0.4988          & 0.6705          & 0.4992          & \multicolumn{1}{c}{0.6716}          & 0.5175          & 0.6765          & 0.6772          & 0.6016           \\
$\checkmark$    & $\checkmark$    & $\times$        & 0.5018          & 0.6682          & \cellcolor{greyL}\textbf{0.5038}          & \multicolumn{1}{c}{0.6729} & 0.5126          & 0.6746          & 0.6735 & 0.6010           \\
$\checkmark$    & $\checkmark$    & $\checkmark$    & \cellcolor{greyL}\textbf{0.5083} & \cellcolor{greyL}\textbf{0.6808}          & 0.5029 & \cellcolor{greyL}\textbf{0.6787}                      & \cellcolor{greyL}\textbf{0.5194} & \cellcolor{greyL}\textbf{0.6841} & \cellcolor{greyL}\textbf{0.6879} & \cellcolor{greyL}\textbf{0.6089}  \\
\bottomrule[1.2pt] 
\end{tabular}}}
\end{table*}

\begin{figure*}[!t]
 \centering
    \includegraphics[width=0.98\textwidth]{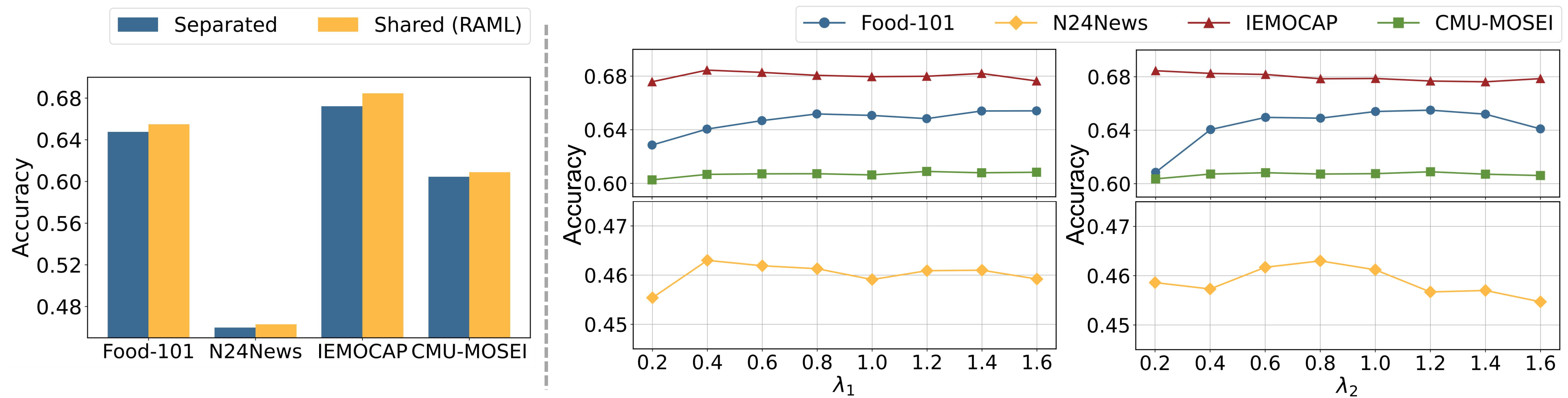}
    \caption{Left: The comparison of RAML and RAML with separate predictors on four datasets. Right: The performance of RAML under different $\lambda_1$ and $\lambda_2$.}
    \label{fig6}
\end{figure*}


\begin{figure*}[!t]
 \centering
 \includegraphics[width=1.0\textwidth]{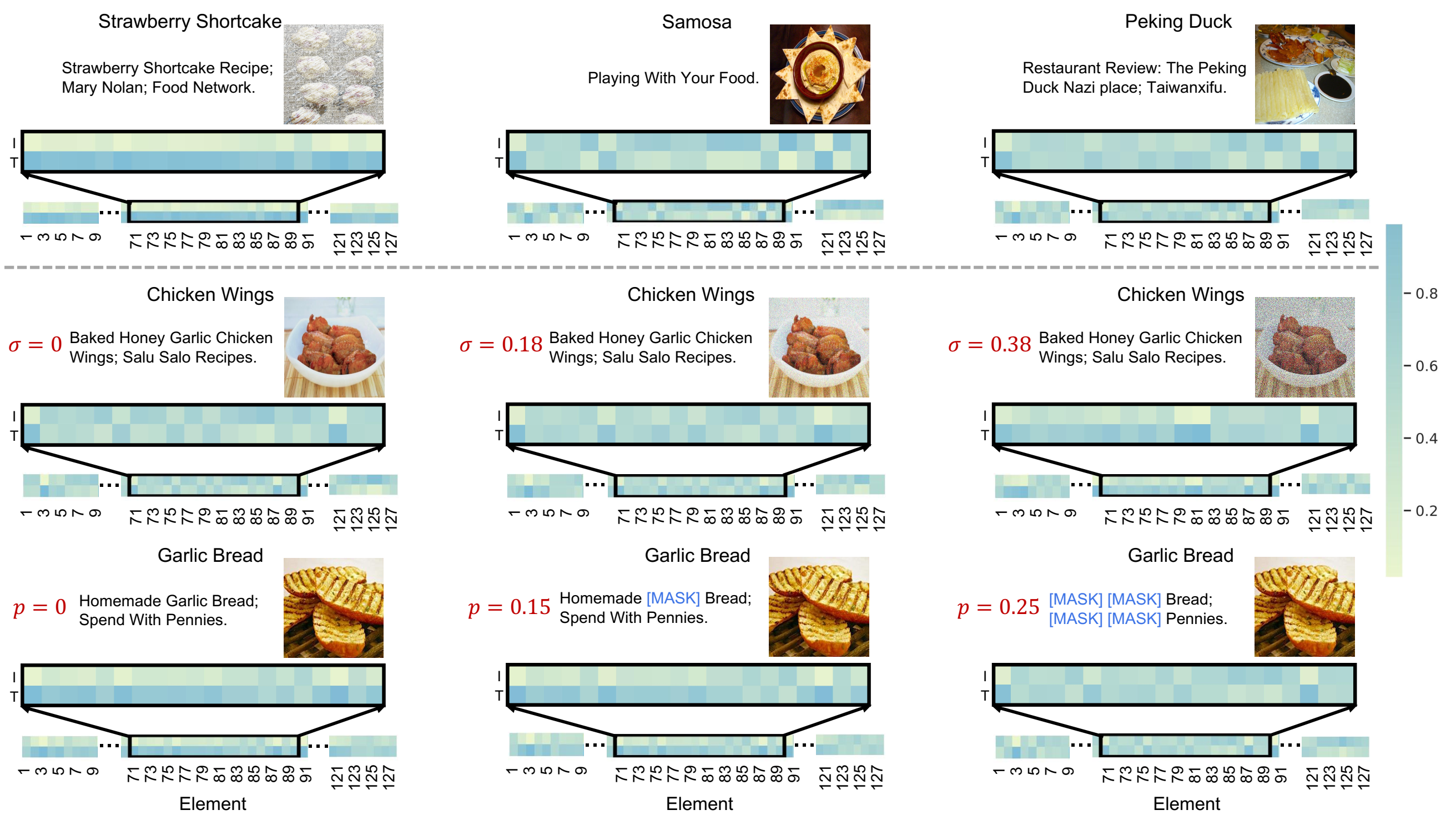}
    \caption{Top: Heatmap of the element-wise weights of the text and image modalities from different samples on Food-101. The weights change over the quality of different data.
    Bottom: Heatmap of weights varying with changing Gaussian noise intensity on an image or changing masking word probability on a text. Here, the variance values $\sigma$ of the noise is set as 0.0, 0.18, 0.38 and the masking probability $p$ as 0, 0.15, 0.25 separately. In general, the element-wise weights of the image decrease as noise intensity increases, while the weights of the text decline with higher masking probabilities.}
    \label{fig5}
\end{figure*}

\subsection{Further Analysis}

\noindent\textbf{Ablation on Adaptive Weights.} To verify the effectiveness of dynamically learning element-wise unimodal weights for different samples, we conducted experiments on two variants of RAML: using identical weights for each available modality or using different but fixed element-wise weights. The results are shown in Table~\ref{tab:weight}. We can see that using fixed element-wise weights is better than using identical weights, but worse than learning adaptive weights. For example, the performance of using fixed element-wise weights is 0.83\% higher on average than using identical weights but 1.23\% lower than RAML on IEMOCAP. This phenomenon indicates that the fine-grained quality of information from each modality changes with different elements and samples. Learning these varying qualities promotes the fusion of available modalities.
\vspace{-1pt}

\noindent\textbf{Effectiveness of the Shared Predictor. }To verify the effectiveness of the shared predictor, we compare the performance of RAML to that without the shared predictor. The results are shown in the left half of Figure~\ref{fig6}. 
We can see that training RAML with separate predictors results in some performance degradation. Specifically, without the shared predictor, the average results of all missing-modality cases decrease by 0.64\%, 0.2\%, 1.1\%, 0.55\% on average on Food-101, N24News, IEMOCAP and CMU-MOSEI respectively. The shared predictor ensures the semantic coherence of multimodal and unimodal embeddings, which not only makes unimodal embeddings comparable but also further boosts the elimination of spurious information in single modalities.
\vspace{-1pt}

\noindent\textbf{Ablation on Loss Components.} In this part, we study the impact of each loss component in Eq.~\eqref{eq:all} to show their effectiveness in improving the multimodal robustness on Food-101, N24News, IEMOCAP and CMU-MOSEI. 
In Table~\ref{tab:3}, we conduct the ablation study and summarize the corresponding performance with or without different loss components. According to the results in Table~\ref{tab:3}, we have the following observations: 
1) The model using the unimodal and multimodal classification losses $\mathcal{L}^U, \mathcal{L}^M$ performs better than the simple model with only $\mathcal{L}^M$. Additionally, RAML outperforms the model using $\mathcal{L}^M$ and $\mathcal{L}^D$. For instance, in terms of the "Average" metric on Food-101, the performance drops from 65.40\% to 57.59\% when optimizing RAML without $\mathcal{L}^U$. Similarly, on N24News, the performance drops from 46.30\% to 44.59\%. These results suggest that constraining the unimodal embedding to be discriminative indeed helps to model the more informative unimodal distributions and to have a better fusion representation. 2) The model with both $\mathcal{L}^M$ and $\mathcal{L}^D$ outperforms the simple model. Additionally, RAML with all loss components outperforms the model with $\mathcal{L}^M$ and $\mathcal{L}^U$. For example, in comparison, RAML achieves a 2.59\% and 0.79\% improvement on average over the model with $\mathcal{L}^M$ and $\mathcal{L}^U$ on Food-101 and IEMOCAP, respectively. This performance gain validates the effectiveness of the proposed loss $\mathcal{L}^D$ in eliminating spurious information, which is helpful for perceiving the untainted unimodal information. 

\noindent \textbf{Hyperparameter $\lambda_1$ and $\lambda_2$.} In the right half of Figure~\ref{fig6}, we conducted several experiments with different values of $\lambda_1$ and $\lambda_2$ separately to validate the stability of RAML. We compared values of $\lambda_1$ and $\lambda_2$ ranging from 0.2 to 1.6. From the curve, we can see that setting a small value of $\lambda_1$ makes it difficult for the model to capture useful unimodal information for the task, thus affecting the final performance of multimodal fusion. Additionally, $\lambda_1$ appears to be insensitive within a certain range (0.4-1.4). The hyperparameter $\lambda_2$ controls the sparsity of the unimodal embeddings and should be properly chosen; setting it to a small value does not effectively eliminate spurious information, and a large value hurts the unimodal representation learning. 

\noindent \textbf{The Element-wise Measure under Different Samples.} We take the text and image embeddings of different samples with changing quality from the Food-101 dataset as examples to illustrate and visualize the element-wise weights (\ie $(\frac{\sigma}{\sigma_m})^2$ in Eq.(\ref{eq2})) in the top half of Figure~\ref{fig5}. From the heatmap, we can observe that RAML could perceive the information quality of unimodal embeddings for different samples. To be specific, RAML assigns a larger weight to the text embedding when the image quality is poor. Conversely, the image embedding is assigned a greater weight when the semantics of the text are independent of the ground-truth label. 

\noindent \textbf{The Element-wise Measure under Corruption.} We also visualized the weight changes of the two modalities on Food-101 when one modality encounters different levels of noise. In the second row of Figure~\ref{fig5}, Gaussian noise with variances of 0, 0.18 and 0.38 is applied to the image modality, and in the last row, the masking probability of the text is 0, 0.15 and 0.25.
In general, it can be seen that the element-wise weight of the image modality decreases as the intensity of Gaussian noise on the picture increases, while the weight of the text modality decreases with the rising number of masking words. 
These visualization results demonstrate that our proposed RAML successfully captures the changing quality of unimodal information by generating dynamic weights, which promotes multimodal fusion to become more flexible and efficient.

\section{Conclusion}

In this paper, we introduce the RAML method to handle the problem of data imperfection via leveraging the redundancy and complementarity among modalities. To aggregate multimodal information effectively and efficiently, RAML learns unimodal representations that fully capture useful information including both the redundant and complementary parts, and measures the fine-grained quality of each unimodal representation adaptively. Our experiments demonstrate that RAML can achieve state-of-the-art performance under various imperfect conditions using a simple and intuitive architecture. We hope RAML will serve as a solid baseline and help the future research of multimodal robustness.


\appendix[Proof of Theorem 3.1]
\noindent\textit{Proof.\quad}Ideally, assuming that the non-uniform weights in Eq.(\ref{eq:power}) are optimal for recognizing the pattern of the unimodal representations $\mu_1^i,\mu_2^i$, which means that the larger one between $\sum_d \theta^T_d \mu_{1,d}^i$ and $\sum_d \theta^T_d\mu_{2,d}^i$ in $P_c$ or between $\theta^T_d \mu_{1,d}$ and $\theta^T_d\mu_{2,d}^i$ in $P_f$ is assigned to the weight of 1, while the smaller one is assigned 0. Thus, we can express the upper bounds of Eq.(\ref{eq:power}) as follows,
\begin{align}
   & \sup P_u \! = \! \frac{1}{1  +  \mathrm{exp}\left(- \left(\frac{1}{2} \sum_d \theta^T_d \mu_{1,d}^i + \frac{1}{2} \sum_d \theta^T_d\mu_{2,d}^i\right) \right)}, \nonumber\\
   & \sup_\alpha P_c \! =  \! \frac{1}{1  +  \mathrm{exp}\left(- \max \! \left(\sum_d \theta^T_d \mu_{1,d}^i, \sum_d \theta^T_d\mu_{2,d}^i\right) \right)},  \\ 
   & \sup_\omega P_f \! = \! \frac{1}{1  +  \mathrm{exp}\left(- 
   \sum_d \max \! \left( \theta^T_d \mu_{1,d}^i,  \theta^T_d\mu_{2,d}^i \right) \right) }. \nonumber 
\end{align}
Since the inequality relationship
$\frac{1}{2} \! \sum_d \! \theta^T_d \mu_{1,d}^i  +  \frac{1}{2} \! \sum_d  \! \theta^T_d\mu_{2,d}^i \leq  \max \! \left(\sum_d \! \theta^T_d \mu_{1,d}^i, \sum_d \! \theta^T_d\mu_{2,d}^i\right) \leq \sum_d \max \! \left( \theta^T_d \mu_{1,d}^i,  \theta^T_d\mu_{2,d}^i \right)$ always holds true, we can get that $\sup_\omega P_f\geq \sup_\alpha P_c \geq \sup P_u$.

\bibliographystyle{plain} 
\bibliography{sample-base}


 




\vfill

\end{document}